\documentclass[11pt]{article}

\newcommand{\blind}{0}

\usepackage{amsmath}
\usepackage{graphicx}
\usepackage{enumerate}
\usepackage{enumitem}
\usepackage{natbib} 
\usepackage{url} 

\graphicspath{{figures/}}
\usepackage{amsfonts, amsthm, latexsym, amssymb}
\usepackage{dsfont}
\usepackage{array}
\usepackage{float}
\usepackage{tikz}
\usepackage{multirow}
\usepackage{xcolor}
\usepackage{lineno}
\usepackage{pgfplots}
\pgfplotsset{compat=1.17}
\usepackage{hyperref}

\usepackage{authblk} 
\usepackage{booktabs}
\usepackage[utf8]{inputenc}
\usepackage[english]{babel}
\usepackage{csquotes} 
\usepackage{graphicx}
\usepackage{subcaption}
\usepackage{algpseudocode}
\usepackage{tikz}
\usetikzlibrary{positioning, arrows.meta}
\usepackage[ruled,vlined]{algorithm2e}

\begin{document}

\def\spacingset#1{\renewcommand{\baselinestretch}%
{#1}\small\normalsize} \spacingset{1}

		\if0\blind
  {
\title{Domain-Adapted Power Curve for Cross-Farm Applications}

\newcommand{\authorlastnames}{Chokhachian, Joseph, Ding}
\author{Ahmadreza Chokhachian, V. Roshan Joseph and Yu Ding\thanks{Corresponding author: Yu Ding, yu.ding@isye.gatech.edu.} \\
H. Milton Stewart School of Industrial and Systems Engineering\\
Georgia Institute of Technology
}

}\fi
		\if1\blind
		{
\title{Domain-Adapted Power Curve for Cross-Farm Applications}
			\author{Author information is purposely removed for double-blind review}
   \bigskip
			\bigskip
			\bigskip
}\fi

\date{\small \today}
\maketitle
\begin{abstract}
The wind energy industry relies on accurate power curve models to make power forecast, evaluate turbine performance, quantify upgrade, or support site-planning decisions. In this paper, we focus on site-planning power curves, i.e., we investigate how power curve models trained using turbine data on an operating wind farm can be transferred to a new, undeveloped farm. The traditional wisdom in the wind energy literature relies on distance, layout, or terrain characteristics for making cross-farm power curve transfer. Through the lens of domain adaptation, we propose a more reliable transfer learning approach for cross-farm power curve modeling. In the cross-farm applications, a domain is specified by the temporal environmental variates and spatial terrain variables. Domain adaptation is to find a capable similarity metric to adapt the domain on the new farm to that on the existing farm. Empirical results show that our domain adapted power curve consistently outperforms competing approaches by an appreciable margin for site-planning power predictions.
\end{abstract}

\noindent%

\spacingset{1.5}

\section{Introduction}

One of the most fundamental functions in wind turbine analysis is the wind turbine power curve \citep{Ding2019}, which is a functional relationship, \( y=f(\mathbf x) \), connecting a turbine's power output $y$ with the environmental conditions in $\mathbf x$ under which the turbine is operating. Among the elements in \( \mathbf{x} \), wind speed is understandably the most influential variable, but previous studies revealed that additional inputs such as wind direction, temperature, turbulence intensity, and even humidity may also play certain roles. The main purpose of the power curve is for wind power forecasting, which can be achieved by mapping the forecasted environmental conditions through the power curve. It can also be used for turbine performance assessment~\citep{Ding2022}, where deviations from the expected power curve can indicate efficiency losses, degradation, or abnormal operations \citep{Lee2015b}.

Estimating a wind turbine power curve is inherently a nonparametric regression problem. The International Electrotechnical Commission (IEC)'s \emph{binning} method is one of the simple nonparametric regression approaches. The IEC binning method estimates wind power output as a function of wind speed by averaging observed power within fine wind speed intervals \citep{IEC2005}. Researchers realized that machine learning methods can be used to build power curves with better performance, and indeed, many machine learning-based power curve models have been developed over the past decade, which was comprehensively discussed in Chapter 5 of \citet{Ding2019} and implemented through its companion packages in R \citep{Kumar2020} and Python \citep{Kumar2022}, including a kernel method \citep{Lee2015a}, k-NN \citep{Yesilbudak2013}, support vector machine \citep{Pandit2020}, trees \citep{Kusiak2009}, smoothing splines \citep{Gu2013}, deep neural network \citep{Wang2023} and Gaussian process \citep{Prakash2023}. Data challenges were organized to investigate the most competent power curve functions \citep{Barber2024}.

The aforementioned power curve models are \emph{turbine specific} or \emph{site specific}. What this means is that the existing power curves are learned using the historical wind and power measurements obtained on a specific turbine at a fixed site. Because the data used in learning encodes local wind and operating conditions, the turbine-specific power curves characterize the relationship between wind features and power output for those individual turbines and are better suited for \emph{post-installation applications} associated with specific turbines; see the left-top panel of Figure~\ref{fig:fig1}.

There are needs going beyond post-installation, turbine-specific applications.  One of such circumstances is wind farm planning when decision makers would like to decide which turbines to purchase, where to install them, and even, how to position them on a new wind farm.  For those types of decisions, it is crucial to have the power curve of wind turbines on the new farm under planning. One could measure the actual wind and environmental conditions in $\mathbf x$ on the new farm, but because wind turbines have not yet been installed, there are no corresponding wind power output that can be used to learn the power curve for the same type of turbines on the new site. As a result, the turbine-specific approaches do not apply; see the left-bottom panel of Figure~\ref{fig:fig1}. Farm planners may still take the power curve learned on the old wind farm and nevertheless use it on the new farm. Past studies have pointed out ineffectiveness in doing so. For example, \citet{Hammer2022} investigated cross-farm power curve transferability by using the averaged power curves from turbines on the old farm and found that prediction performance does not improve systematically with spatial closeness or it can even deteriorate across nearby sites. In this paper, we ask the question whether there is a better approach to learn a cross-farm power curve using the power data obtained on the old site and the wind data obtained on both old and new sites, i.e. could we transfer the power curve across wind farms? See the right panel of Figure~\ref{fig:fig1}.   

\begin{figure}[h]
\centering
\includegraphics[width=1\linewidth]{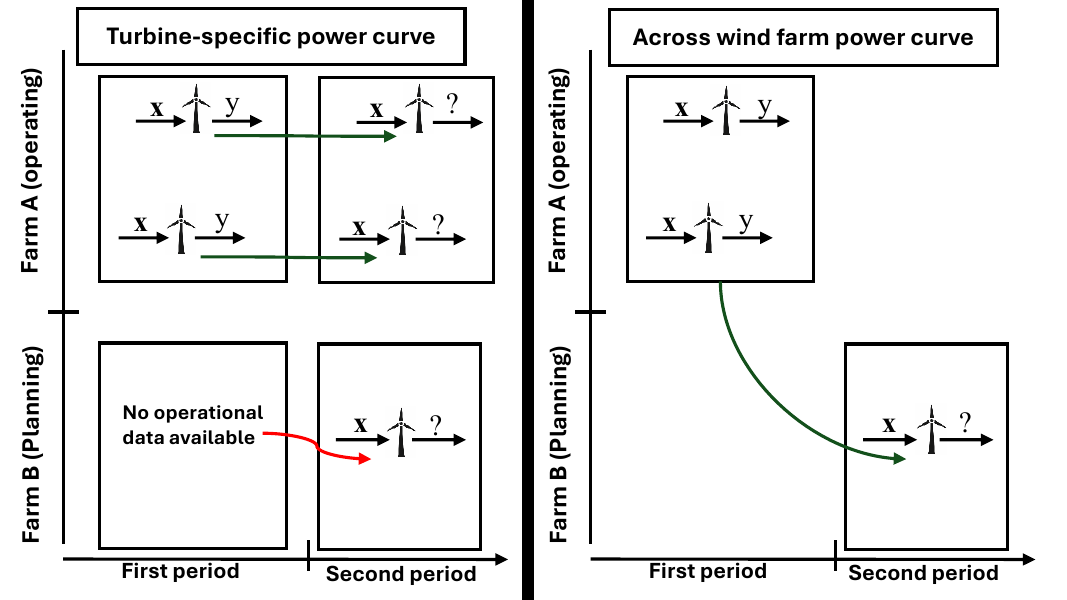}
\caption{Conceptual illustration of turbine-specific power curves (left panel) and cross–farm power curves (right panel).}
\end{figure}\label{fig:fig1}

There has been research addressing the question concerning cross-farm transfer of power curves.  The key is how to account for the difference between the new site and the old site. The existing literature falls into three schools of thought, each attributing successful transfer to one of three key factors: geographical distances, layout structure, and terrain.

The first school does not explicitly account for the between-site differences, but instead examines \emph{ex post} how transfer performance relates to geographic proximity once models are trained and applied across sites. For example, \citet{li2022b} developed support vector regression models on one wind farm and applied them to other farms but concluded that effective transfer occurs only under highly restrictive conditions, such as when latitude differences are below $0.2^\circ$, inter-farm distances under 50~km, and relatively flat terrain. \citet{Arrieta2024} developed a random forest model for cross-farm prediction and analyzed the drivers of successful transfer. Their results suggest that in addition to geographic similarity (such as latitude), similarity in turbine layout plays a role, too. In this category of approaches, geography is used to interpret the outcomes rather than as a modeling input. 

The second school of power-curve transfer leverages geographic and layout structure \emph{implicitly} through the model architecture. In this line of work, \citet{Daenens2025} proposes a Graph Neural Network (GNN) in which each wind farm is represented as a graph, with the nodes defined by the turbines and the edges defined by the inter-turbine distances and relative orientations. Time-series measurements at each turbine are modeled using an LSTM (Long Short-Term Memory) encoder, whereas the edge features describing the wind farm layout are encoded via an MLP (Multi-Layer Perceptron). These node and edge representations are combined through message-passing layers to learn shared latent representations across turbines and wind farms. When applied to a geographically distant or previously unseen farm, the GNN model primarily relies on the node-level temporal representations, making the physical drivers of a potentially successful transfer difficult to interpret explicitly. Implicitly, one could credit geographic distances or similarity in turbine layouts when a transfer goes well. 

The third school seeks to account for site differences by explicitly modeling similarities in terrain characteristics. To the best of our knowledge, \citet{Prakash2024} and \citet{chokhachian2026} are the two power curve models that make explicit use of terrain characteristics as part of input in their respective models. \citet{Prakash2024} developed a Bayesian  hierarchical model (BHM), whereas \citet{chokhachian2026} developed a spatial-temporal Gaussian process model (STGP). \citet{Prakash2024} and \citet{chokhachian2026} showed that by taking the terrain information at the new site as part of the input, their model can borrow strength across sites and enable cross-farm power curve transfer. Because the approaches in \citet{Prakash2024} and \citet{chokhachian2026} rely on terrain measures for transfer and not on geographic distances, their transferability goes beyond the same or nearby farms.

In this paper, we view the cross–farm power curve modeling through the lens of domain adaptation. Domain adaptation is one form of transfer learning \citep{PanYang2010} and focuses on adapting a model trained on a source domain for effective application in a target domain where there are differences between the source and target domains (the so-called domain shift). In our problem setting, a turbine on the existing wind farm provides a source domain, whereas the target turbine on the new site defines the target domain. Domain adaptation enables model transfer by identifying source data that are most relevant to the target setting. In fact, the existing cross-farm power curve transfer papers discussed above made an \emph{implicit} use of the domain adaptation idea in the sense that their approach assumed that either the geographic proximity, or the similarity in farm layout, or terrain similarity is sufficient to ensure a good adaptation. In practice, power curve behavior is more complicated than what can be fully characterized by  site characteristics.

Instead of relying on site characteristics such as distance, layout, or terrain alone, we believe that a more principled approach is to characterize the source/target domain using the combination of a turbine's temporal environmental covariates and spatial terrain characteristics. Moreover, we develop a supervised distributional similarity metric that effectively aligns the source and target domains. Using this metric, we identify the subset of turbines whose temporal covariates and terrain characteristics most closely resemble those of the target turbine. Aggregating the training data from these selected turbines we construct the cross-farm power curve. Our numerical analysis is conducted on a wind farm comprising 66 turbines across varying degrees of terrain complexity, and demonstrates that the proposed cross-farm power curve outperforms the competing alternatives including those relying on site characteristics for power curve transfer.

The remainder of the paper is organized as follows. Section~\ref{sec:data} describes the dataset used in this study. Section~\ref{sec:model} then presents the proposed domain adaptation framework. Section \ref{sec:experiments} evaluates the proposed method and compare its performance with the alternative methods. Finally, Section~\ref{sec:conclusion} concludes the paper with a discussion on future research.

\section{Data Description}
\label{sec:data}

\begin{figure}[h]
\centering
\includegraphics[width=0.8\linewidth]{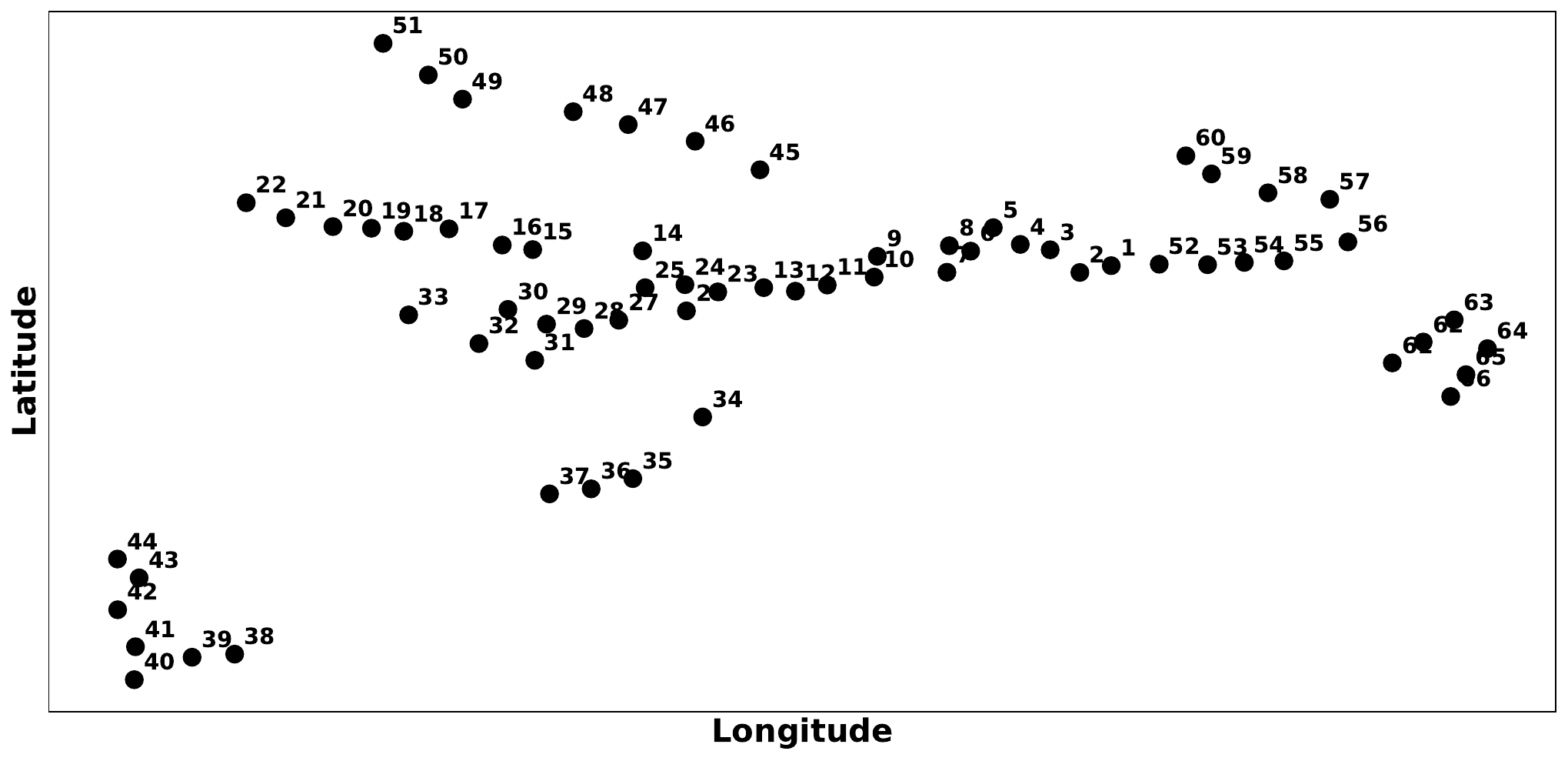}
\caption{Wind farm layout}
\label{fig:fig5}
\end{figure}

The dataset used in this study comprises 66 turbines from an onshore wind farm. The layout of the wind farm is shown in Figure~\ref{fig:fig5}. The wind farm consists of turbines located in regions with varying terrain characteristics and elevations. Although the dataset spans 2014 to 2019, the years 2017 and 2018 have the fewest missing data points. Therefore, in this study, we treat 2017 as the first period and 2018 as the second period.

The original data were recorded at a higher frequency and then averaged to 10-minute intervals to align with the standard SCADA (Supervisory Control and Data Acquisition) resolution. For each year, a complete dataset would contain 52,560 observations per turbine. However, after preprocessing, approximately 40,000 to 45,000 observations remain for each turbine, depending on the extent of missingness and the duration of turbine downtime.

The covariates include nacelle-measured wind speed $V$ (m/s), nacelle wind direction $D$ (degrees, $0\text{--}360$), ambient temperature $T$ ($^\circ$C), turbulence intensity $TI$ (defined as the ratio of the standard deviation of wind speed to its mean), and the standard deviation of wind direction $sdD$. The response variable is turbine power output, scaled to the interval $[0,100]$, denoted by $y$.

We define the environmental covariate vector $\mathbf{x}$ as
\[
\mathbf{x} = (V, T, D, TI, sdD)^\top.
\]
We index training turbines by $i \in \{1,\ldots,m\}$, where $m=66$ in our dataset. For turbine $i$ at time $t$, the $5 \times 1$ vector of environmental covariates is denoted by
\[
\mathbf{x}^{(i)}_t
= \big(x^{(i)}_{1,t}, \ldots, x^{(i)}_{5,t}\big)^\top,
\qquad t = 1,\ldots,n_i,
\]
where $n_i$ is the number of available observations for turbine $i$ and the elements in $\mathbf{x}^{(i)}_t$ takes the physical meaning of the corresponding element in $\mathbf{x}$, e.g., $x^{(i)}_{1,t}$ is wind speed $V$, $x^{(i)}_{2,t}$ is wind direction $D$, and so forth.  Understandably, in other wind applications, the number of elements in $\mathbf x$ may be more than five, but the notation can be easily extended to accommodate more environmental covariates.

Collecting all covariates for turbine $i$ gives
\[
\mathbf{X}^{(i)} =
\begin{pmatrix}
\,| & | & \cdots & |\, \\
\mathbf{x}^{(i)}_1 & \mathbf{x}^{(i)}_2 & \cdots & \mathbf{x}^{(i)}_{n_i} \\
\,| & | & \cdots & |\,
\end{pmatrix}
\in \mathbb{R}^{l \times n_i},
\]
where each column corresponds to one observation $\mathbf{x}^{(i)}_t$. The corresponding response vector is
\[
\mathbf{y}^{(i)} =
\big(y^{(i)}_1,\ldots,y^{(i)}_{n_i}\big)^\top
\in \mathbb{R}^{n_i},
\]
where $y^{(i)}_t$ denotes the power output at time $t$. Both the covariates $\mathbf{x}^{(i)}_t$ and the response $y^{(i)}_t$ vary across time and turbines. 

In addition to the environmental covariates, terrain measurements are available for each turbine in the wind farm. Following the IEC 61400-12-2 standard \citep{IEC2013}, the physical characteristics of the terrain surrounding each turbine are described through three variables: the inclination of the terrain surface measured in degrees (slope), the fraction of the surrounding terrain exceeding a critical steepness threshold (ruggedness index, RIX), and the elevation of the highest nearby ridge in meters. Each of these descriptors is evaluated across 36 directional sectors of $10^\circ$ each, yielding a $36 \times 3$ matrix as the most granular representation of terrain at a given turbine. Aggregating across sectors produces a $3 \times 1$ summary vector, and further compression assigns each turbine to one of five discrete terrain complexity categories, ranging from 1 (flat) to 5 (highly complex), according to the classification rules specified in \cite{IEC2013}.

Rather than aggregating the sector-wise measurements through a simple average, we follow \citet{Prakash2024} and compute a wind-direction-weighted average, which assigns greater weight to sectors through which wind flows more frequently. Formally, let $\mathbf{s}_{b,i}$ denote the $3 \times 1$ terrain descriptor vector for direction sector $b$ at turbine $i$, and let $\gamma_{b,i}$ be the empirical frequency of wind in that sector, with $\sum_{b=1}^{36} \gamma_{b,i} = 1$. The resulting terrain representation for turbine $i$ is then
\begin{equation}
    \mathbf{s}_i = \sum_{b=1}^{36} \gamma_{b,i} \, \mathbf{s}_{b,i},
    \label{eq:terrain}
\end{equation}
which captures the effective terrain exposure of each turbine as experienced under its actual wind conditions. Unlike environmental covariates, terrain features are static. They vary across turbines but remain constant over time for any given turbine. Environmental covariates and power output, by contrast, vary both across turbines and over time.

Throughout this study, our main evaluation strategy is leave-one-turbine-out (LOTO) cross-validation. In LOTO, one turbine is held out for testing while the remaining turbines are used for training; this process is repeated for all 66 turbines, and the final error is computed as the average across all test turbines. By averaging over all turbines, LOTO provides a stable error estimate and is therefore well-suited for guiding design choices such as the aggregation strategy and the number of source turbines.

\section{Model}\label{sec:model}

The central challenge addressed in this work is the construction of a predictive power curve for a target turbine that has no historical power observations. A commonly used approach is to model the relationship between covariates and power output using all available turbines of same model and make on an operating farm:
\begin{equation}\label{eq:model1}
y_t^{(i)} = f\!\left(\mathbf{x}_t^{(i)}, \mathbf{s}_i\right) + \epsilon_t^{(i)},
\qquad i = 1,\ldots,m,\;\; t = 1,\ldots,n_i,
\end{equation}
where $f(\cdot)$ represents a global power curve model that depends 
on both the time-varying environmental covariates $\mathbf{x}_t^{(i)}$ 
and the static terrain descriptor $\mathbf{s}_i$, and $\epsilon_t^{(i)}$ 
is a noise term. The goal is to estimate $f$ using data from existing 
turbines and then use the resulting $\hat{f}$ to predict power 
output at a target location on the planning farm by plugging in $\mathbf{x}_t^{\mathrm{target}}$ 
and $\mathbf{s}_{\mathrm{target}}$, which are, respectively, the temporal covariates and terrain measurements at the target location.

Turbines on a wind farm do operate under diverse environmental conditions. Some of them may be closer to that of the target turbine while others are less so. Fitting a single global model as in \eqref{eq:model1} is to use the average turbine on an existing farm to represent any target turbine on a new farm. Using the language in domain adaptation, when the source and target distributions differ, a model trained indiscriminately on all sources may fail to capture the conditions most relevant to the target. 

To address this, we propose a two-step approach. The first step is to adapt the domain, which is to identify a subset of turbines whose operating conditions are most similar to the target. Here we define our domain using the combination of the environmental variates in $\mathbf x$ and terrain descriptors in $\mathbf s$, i.e., $(\mathbf x, \mathbf s)$. Then, the domain adaptation is to find the set of $P \subset \{1,\ldots,m\}$, which selects the $K$ turbine indices that have the closest domain matching, i.e.,
\begin{equation}\label{eq:distance}
P = \operatorname{arg\,min}_{i \in \{1,\ldots,m\}}^{(K)}
d\!\left[\left(\mathbf{x}^{(i)}, \mathbf{s}_i\right),\, 
\left(\mathbf{x}^{\mathrm{target}}, \mathbf{s}_{\mathrm{target}}\right)\right],
\end{equation}
where the superscript $(K)$ means that the minimization operation selects the $K$ smallest values and $d[\cdot,\cdot]$ is a dissimilarity measure between covariates, to be specified in Section~\ref{subsec:matching}. 

The second step is to train a power curve model using the data from the chosen turbine subset:
\begin{equation}\label{eq:model2}
y_t^{(j)} = f\!\left(\mathbf{x}_t^{(j)}, \mathbf{s}_j\right) + \epsilon_t^{(j)},
\qquad j \in P,\;\; t = 1,\ldots,n_j.
\end{equation}
This is to say, $\hat{f}$ is trained using only observations from 
turbines in $P$ and then used for making power prediction at a target location with 
$\mathbf{x}_t^{\mathrm{target}}$ and $\mathbf{s}_{\mathrm{target}}$.

\subsection{Source Turbine Selection}\label{subsec:matching}
In the language of domain adaptation, the operating conditions of a turbine is the domain, and the environmental covariates $\mathbf x$ and terrain variables $\mathbf s$ used to characterize a turbine's operating conditions are the domain characteristics. Domain adaptation, i.e., the action to adapt the target domain to the source domain, is achieved via source matching, which is, in our problem, to select the source turbines whose domain characteristics, i.e., the operation conditions, to match those of the target turbines.  For the purpose of enhancing robustness, $K$ source turbines, instead of a single one, are selected through source matching, where $K$ is a parameter set \emph{a priori}. 

For such a type of matching and selection, one needs a dissimilarity measure, which is a distance metric. Because the variables in $\mathbf x$ and $\mathbf s$ behave differently, we decided to choose different dissimilarity measures to compare them.

The variables in the terrain descriptor $\mathbf s$ are a single scalar value per turbine because these values do not change with time. We believe it is sufficient to use the absolute difference to quantify the difference between a pair of measurements. That is, for terrain covariate $c \in \mathcal{C}_s$, we first min-max scale $s^{(c)}$ to the unit interval $[0, 1]$, and the per-covariate dissimilarity is then defined as:
\begin{equation}
d^{(s)}_c(i,\mathrm{target}) =
\bigl| s^{(c)}_i - s^{(c)}_{\mathrm{target}} \bigr|.
\end{equation}

The temporal environmental covariates such as wind speed are observed over a period of time and are highly stochastic. We choose to use a probability distribution to capture its information.  As such, for environmental covariate $c \in \mathcal{C}_x$, let $F^{(c)}_i$ and $F^{(c)}_{\mathrm{target}}$ denote the empirical cumulative distribution functions based on marginal samples
\begin{equation}
\mathbf{X}^{(i)}_{c} = \{ x^{(i)}_{c,t} : t = 1, \ldots, n_i \},
\qquad
\mathbf{X}^{(\mathrm{target})}_{c} = \{ x^{(\mathrm{target})}_{c,t}
: t = 1, \ldots, n_{\mathrm{target}} \}.
\end{equation}
We use the Kolmogorov--Smirnov (KS) distance to quantify the difference between the two measurements of the same covariate $c$ in $\mathbf x$:
\begin{equation}
d^{(x)}_c(i,\mathrm{target}) =
\sup_{x} \bigl| F^{(c)}_i(x) - F^{(c)}_{\mathrm{target}}(x) \bigr|,
\qquad c \in \mathcal{C}_x.
\end{equation}

Then, the two separate covariate-specific dissimilarity measures are aggregated into a unified metric, denoted by $d_{WD}$, which is the weighted dissimilarity metric (WD).
\begin{equation}
d_{WD}\!\left[(\mathbf{X}^{(i)}, \mathbf{s}_i),\,
(\mathbf{X}^{(\mathrm{target})}, \mathbf{s}_{\mathrm{target}})\right]
= \sum_{c \in \mathcal{C}_s} w_c \, d^{(s)}_c(i,\mathrm{target})
+ \sum_{c \in \mathcal{C}_x} w_c \, d^{(x)}_c(i,\mathrm{target}),
\end{equation}
where $\sum_{c \in \mathcal{C}_s \bigcup \mathcal{C}_x} w_c = 1$. 

To compute the weights $w_c$, we need a measure of how much each covariate contributes to predicting power output. Intuitively, if wind speed $V$ has a stronger effect on $y$ than turbulence intensity $TI$, we expect the wind speed to receive a higher weight. To obtain the weights, we rely on accumulated local effects (ALE) of \citet{Apley2016}. ALE computes the effect of each varying feature on the response using a black box model $g$. The \emph{local effect} of covariate $c$ at a point $x_c$ is how much the fitted model's prediction changes as $x_c$ moves slightly, holding the other covariates fixed at values that are actually observed near $x_c$ in the training data. Because it only uses covariate combinations that are actually observed near $x_c$, the local effect never requires extrapolating $g$ into regions with little or no real data.

Formally, let $g(\cdot)$ denote the fitted model, and let $g^{\,c}(x_c, \mathbf{x}_{-c}) \equiv \partial g(x_c, \mathbf{x}_{-c})/\partial x_c$ denote the local effect of $c$ described above. The ALE main effect of covariate $c$ is
\begin{equation}
g_{c,\mathrm{ALE}}(x_c) \equiv \int_{x_{\min,c}}^{x_c} E\big[g^{\,c}(X_c, \mathbf{X}_{-c}) \mid X_c = z_c\big]\, dz_c - \mathrm{constant},
\end{equation}
where the conditional expectation, taken with respect to the conditional density of $\mathbf{X}_{-c}$ given $X_c = z_c$, averages the local effect over the covariate combinations actually observed near $z_c$, and it is integrated over $z_c$ range.

We summarize the importance of covariate $c$ by the range of its estimated ALE main effect,
\begin{equation}
\rho_c = \max_{x_c} \hat{g}_{c,\mathrm{ALE}}(x_c) - \min_{x_c} \hat{g}_{c,\mathrm{ALE}}(x_c).
\end{equation}

Computing ALE ranges requires a fitted model to serve as $g$, which can range from a simple regression to a more complex supervised learning method. Because our covariates include spatially static terrain features alongside dynamic temporal features, we fit $g$ using the spatio-temporal Gaussian process framework of \citet{chokhachian2026}, which is specifically designed to accommodate this combination of feature types. The fitted model uses wind speed, temperature, slope, RIX, and ridge height to predict power output. Table~\ref{tab:ale_weights} reports the resulting ALE main effects for each feature. To obtain weights that sum to 100, we divide each $\rho_c$ by the sum of $\rho$ values across all features. Among the environmental features, only wind speed and temperature receive non-zero weights, with wind speed dominating at $77.7\%$ and temperature contributing just $5.4\%$. Among the terrain features, RIX carries the largest weight, followed by ridge height and slope, with the three terrain features together accounting for roughly $17\%$ of the total weight. This pronounced disparity in weights underscores the importance of not treating all features equally when constructing the dissimilarity metric.

\begin{table}[htbp]
\centering \footnotesize
\caption{Feature importance scores and normalized weights obtained from \texttt{ALE} ranges.}
\label{tab:ale_weights}
\begin{tabular}{lcc}
\toprule
\textbf{Covariate} & \textbf{ALE main effects} & \textbf{Weight (\%)} \\
\midrule
Wind speed    & 94.3  & \textbf{77.74} \\
Temperature   & 6.6   & 5.44 \\
Slope         & 2.6   & 2.14 \\
RIX           & 11.5  & 9.48 \\
Ridge         & 6.3   & 5.19 \\
\midrule
Total         & 121.3 & 100.00 \\
\bottomrule
\end{tabular}
\end{table}

We compare $d_{WD}$ against three alternative distance metrics proposed in transfer learning. These include (1) the mean dissimilarity computes the arithmetic mean of marginal KS statistics across environmental covariates without any feature weighting, which can be seen as the unweighted version of $d_{WD}$, as terrain variables do not matter much in $d_{WD}$; (2) the marginal energy distance \citep{rizzo2016} measures distributional discrepancy using the energy distance computed separately for each covariate and then averaged across all covariates; (3) the Sinkhorn Wasserstein \citep{chizat2020} distance that evaluates similarity through an entropically regularized optimal transport formulation applied to the joint multivariate distribution.

To evaluate the effectiveness of each metric, we conduct an experiment using the LOTO cross-validation. We use the thinned twinGP \citep{chokhachian2025} as the downstream predictive machine learning model in order to get numerical results.  Thinned twinGP is chosen because it was shown as a highly competitive predictive model on autocorrelated data and is arguably the fastest. A method's performance is assessed by its predictive root mean squares error (RMSE) at the target location using a holdout test dataset. Throughout the paper, the ``2017 RMSE” denotes the average RMSE from training on the 2017 data of 65 turbines and testing using the 2017 data of the one left-out turbine, whereas the “2018 RMSE” denotes training on the 2017 data of 65 turbines but testing on the 2018 data of the one left-out turbine. Testing on 2017 data primarily assesses spatial generalization. Testing on 2018 data evaluates both spatial and temporal generalization, since the learned model is applied to a different year on a different turbine. The ``2018 RMSE'' setting reflects more closely the real-world new farm planning scenario as illustrated in Figure~\ref{fig:fig1}.

Table~\ref{tab:metric_comparison} presents the comparisons.  The weighted dissimilarity, i.e., $d_{WD}$, achieves a substantial improvement over the unweighted dissimilarity distance, highlighting the importance of incorporating feature weighting. The Sinkhorn Wasserstein distance performs comparably to $d_{WD}$, whereas the marginal energy distance performs noticeably worse.  

Between the weighted dissimilarity metric and the Sinkhorn Wasserstein distance, we choose the former for a simple reason---computing the Sinkhorn Wasserstein distance requires $O(n^2)$ operations per turbine pair, whereas $d_{WD}$ requires $O(n\log n)$. That translates to a $4{,}200\times$ difference in computation when performed on $n = 45{,}000$ data points, the typical size of one year of SCADA data after cleaning. Imagine the time saving by computing $d_{WD}$ rather than the Sinkhorn Wasserstein distance in LOTO testing for a wind farm with 66 turbines, i.e., 2,145 turbine pairs. 

\begin{table}[htbp]
\centering \footnotesize
\caption{Metric comparison conducted under LOTO setup. Boldface indicates the lowest RMSE in each testing year.}
\label{tab:metric_comparison}
\begin{tabular}{lcccc}
\toprule
\textbf{Metric} & \multicolumn{2}{c}{\textbf{Average 2017 RMSE}} & \multicolumn{2}{c}{\textbf{Average 2018 RMSE}} \\
\cmidrule(lr){2-3} \cmidrule(lr){4-5}
 & \textbf{RMSE} & \textbf{Gap (\%)} & \textbf{RMSE} & \textbf{Gap (\%)} \\
\midrule
Weighted Dissimilarity (WD)     & \textbf{3.34} & 0.00  & \textbf{3.58} & 0.00  \\
Sinkhorn Wasserstein Distance   & 3.35          & 0.30  & 3.59          & 0.28  \\
Mean Dissimilarity (unweighted) & 3.51          & 5.09  & 3.71          & 3.63  \\
Marginal Energy Distance        & 3.81          & 14.07 & 3.95          & 10.34 \\
\bottomrule
\end{tabular}
\end{table}

\subsection{Build the Transferrable Power Curve Model}\label{subsec:modeling}
After choosing the source matching criterion, $d_{WD}$, the next action is to select the best subset of $K$ source turbines in $P = \{p_1,\ldots,p_K\}$ and use their data to build a transferrable power curve model that works effectively for a target location on a new farm. Two design choices arise at this stage: which $K$ to use and how to aggregate across the $K$ selected sources?  Again, we will use the 66-turbine farm data to shed some lights on the choice.

\begin{figure}[h]
\centering
\includegraphics[width=1\linewidth, page=2,
  trim=0 6.5cm 0 0,
  clip]{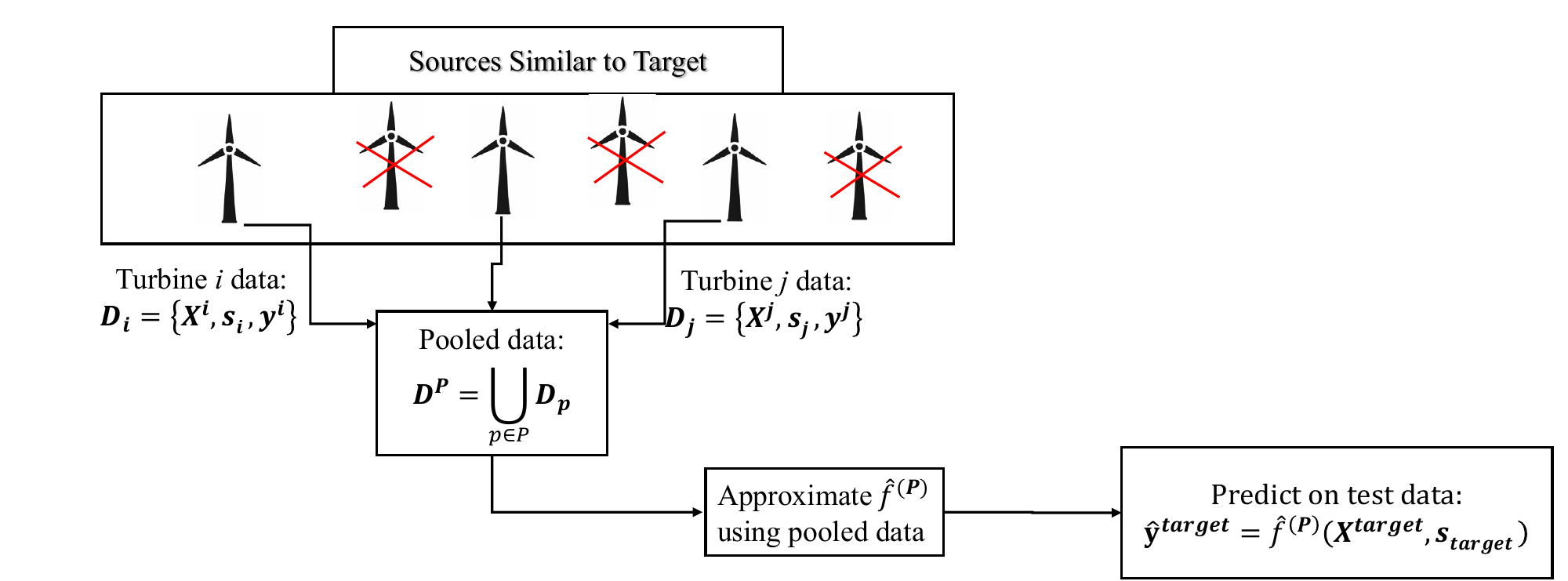}
\caption{Concatenation strategy. Data from all selected source turbines are pooled into a single training set, and a single model is fitted and transferred to the target turbine.}
\label{fig:fig3}
\end{figure}

\begin{figure}[h]
\centering
\includegraphics[width=1\linewidth, page=3,
  trim=0 5cm 0 0,
  clip]{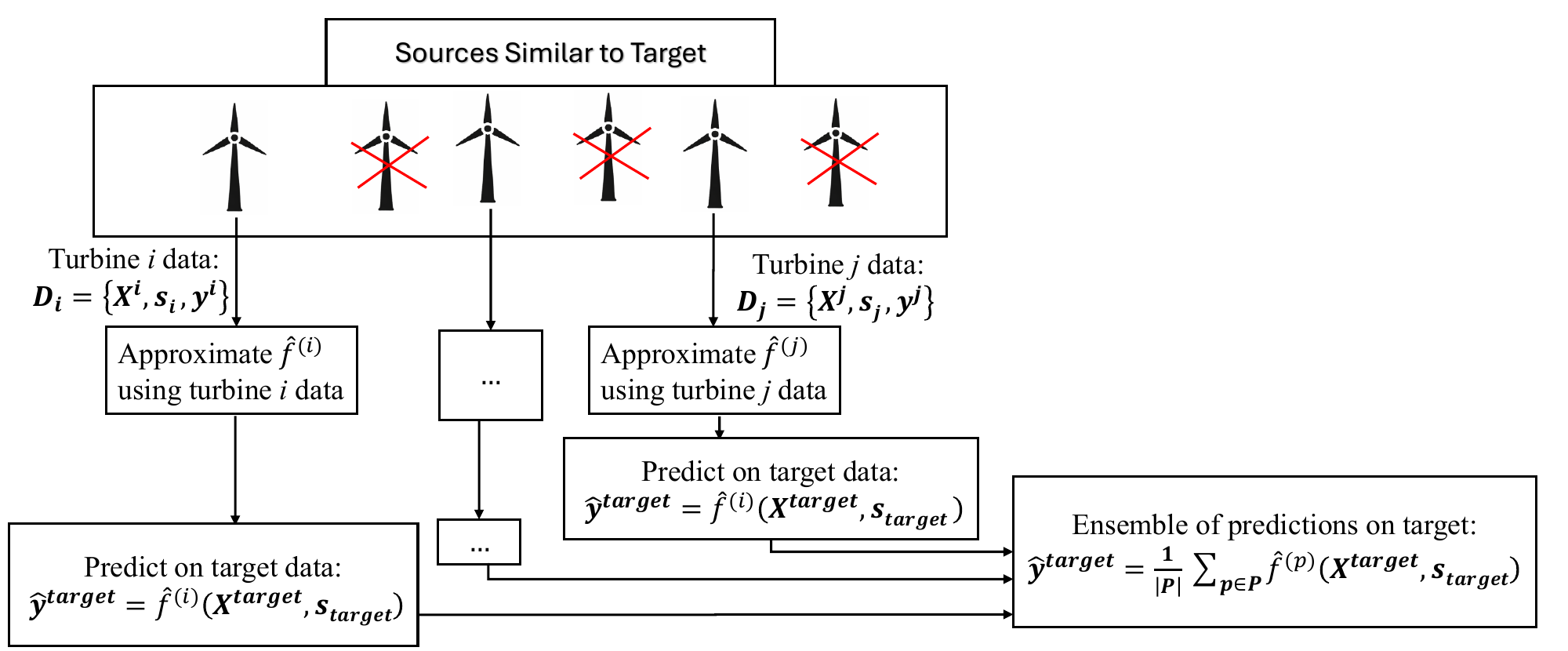}
\caption{Ensemble strategy. A separate model is fitted for each selected source turbine, and predictions are averaged to obtain the final target prediction.}
\label{fig:fig4}
\end{figure}

We consider two aggregation strategies, illustrated in Figures~\ref{fig:fig3} and~\ref{fig:fig4}, respectively.  To explain them, let us first define some notations: let $D_p$ denote the training data for source $p \in P$, $\mathbf{X}^{\mathrm{target}}$ the covariate matrix of environmental features, $\mathbf{s}_{\mathrm{target}}$ the vector of terrain features, and $\hat{\mathbf{y}}^{\mathrm{target}}$ the corresponding predictions at the target. Figure~\ref{fig:fig3} illustrates the concatenation strategy, which pools the data from all selected sources into a single training set,
\begin{equation}
D^{(P)} = \bigcup_{p \in P} D_p,
\end{equation}
trains a single model $\hat{f}^{(P)}$ on $D^{(P)}$, and applies $\hat{f}^{(P)}$ to the target:
\begin{equation}
\hat{\mathbf{y}}^{\mathrm{target}} = 
\hat{f}^{(P)}\!\left(\mathbf{X}^{\mathrm{target}}, 
\mathbf{s}_{\mathrm{target}}\right).
\end{equation}
This approach treats all selected sources as realizations from a shared data-generating process. 

Figure~\ref{fig:fig4} illustrates the ensemble strategy, which fits a separate model $\hat{f}^{(p)}$ to each source $p \in P$ and averages the predictions from all $K$ sources for the target:
\begin{equation}
\hat{\mathbf{y}}^{\mathrm{target}} = \frac{1}{\lvert P \rvert} 
\sum_{p \in P} \hat{f}^{(p)}\!\left(\mathbf{X}^{\mathrm{target}}, 
\mathbf{s}_{\mathrm{target}}\right).
\end{equation}
This strategy preserves heterogeneity among source turbines by allowing each to contribute independently to the final prediction. One natural extension of this strategy is a weighted ensemble, where each source contributes proportionally to its $d_{WD}$ similarity or predictive uncertainty, but numerical experiments show that doing so does not improve much over simple averaging. 

\begin{figure}[htbp]
\centering
\includegraphics[width=1\linewidth]{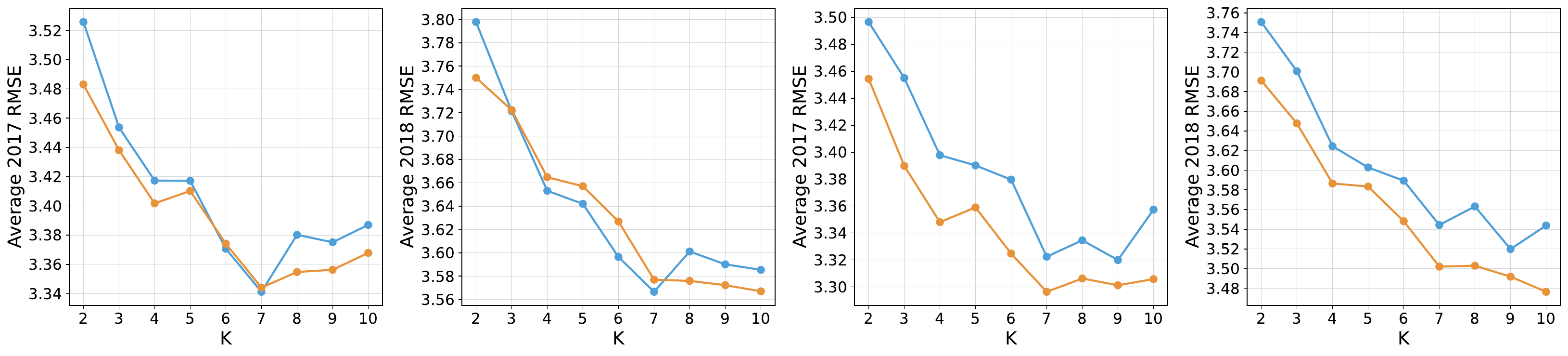}
\caption{Average LOTO RMSE over different $K$'s and either aggregation strategy. The left two plots use the thinned twinGP, whereas the right two plots use a feedforward neural network.  In each panel the orange curve is the ensemble strategy and the blue curve is the concatenation strategy.}
\label{fig:multi_source}
\end{figure}

Again using the data from the 66 turbines but now we use two different machine learning baseline models: one is the thinned twinGP that we used earlier and the other is a multi-layer feedforward neural network \citep{Barber2024}, we test all the combinations of $K$ and the two aggregation strategies. The results of LOTO RMSE, averaged across all 66 turbines, are presented in Figure ~\ref{fig:multi_source}. 
RMSE decreases steadily as $K$ grows from 2 to 7. Beyond that, improvement stalls and becomes inconsistent, with occasional small increases. $K=7$ is a practical sweet spot. For the feedforward neural network and the 2017 RMSE of the thinned twinGP, ensemble averaging consistently outperforms concatenation. For the 2018 RMSE of thinned twinGP, the two strategies are nearly indistinguishable, trading places depending on $K$. Overall, the ensemble strategy is the best choice.

\section{Numerical Experiments}
\label{sec:experiments}

We evaluate the proposed framework on the dataset described in Section~\ref{sec:data}, comprising 66 turbines with 10-minute SCADA 
measurements for 2017 and 2018 and terrain descriptors for all turbines. Some preprocessing and data cleaning were conducted. Transformation is applied to wind direction. Since wind direction is a circular variable, it is replaced by its sine and cosine components. Engineers informed us that four turbines, \# 47, \# 51, \# 53, and \# 61 experienced sensor drift in 2018. Their 2018 records are excluded from evaluation, while their 2017 data are retained for training and testing.

As shown in Table~\ref{tab:metric_comparison}, terrain variables matter much less than the temporal covariates in $\mathbf x$ for this particular wind farm.  So in this numerical study, our resulting power curve model only uses $\mathbf x$ not $\mathbf s$.  But we do not suppose that is always the case.  For some other farms, $\mathbf s$ could play a greater role and a spatio-temporal model may be needed.

In this section, we first evaluate the performance of our method against existing models from the literature under the LOTO cross-validation setting and subsequently focus on a different setup more closely aligned with the farm planning application. We then provide further insights into geographic transfer.

\subsection{Method Comparison on LOTO Cross Validation}

In the LOTO setup, each turbine is treated as the test turbine in turn, while all remaining turbines are used for training. This procedure is repeated for all 66 turbines, and the resulting RMSE values are averaged over the 66 trials. The first comparison group is formed for testing our central assumption that transferring power curves built from a carefully selected subset of turbines yields better predictive performance than using data pooled from all available turbines. Included in this first group are the commonly used power curve methods that do not explicitly model transferability such as XGBoost \citep{XGB}, thinned twinGP \citep{chokhachian2026}, and binning, all of which are designed to scale to large datasets.  What is also included in the first group are the methods reviewed in the earlier section that use some wind farm characteristics to facilitate transferability, including the Bayesian hierarchical method (BHM)~\citep{Prakash2024} and spatio-temporal Gaussian process(STGP) model~\citep{chokhachian2026} that both make use of the terrain descriptors and the GNN-based method~\citep{Daenens2025} that models turbine layout. We used the code provided for public access by BHM and STGP packages for running the numerical comparisons, but had to use our own implementation of \cite{Daenens2025}'s GNN method as their code is not publicly available. 

The second comparison group is formed for testing the hypothesis that the proposed weighted dissimilarity criterion makes a better selection of source turbines than merely the geographic distance. Recall that \citet{Arrieta2024} (random forest), \citet{li2022b} (support vector regression), and \citet{Hammer2022} (XGBoost) emphasize geographic distance as a primary driver of successful transfer, and they used their own choice of a machine learning method (in the parentheses after each citation) to demonstrate that idea. What we want to compare is to apply their choices of machine learning methods (i.e., random forest, XGBoost, or SVR) to a subset of turbines in the training set (instead of all turbines) for model training. But the subset of turbines is selected not using the proposed WD criterion but using the nearest neighborhood criterion, i.e., according to the closest Euclidean between-turbine distance. Concerning the number of turbines in the subset, to maintain consistency, we use the same number as used for the proposed WD transfer, i.e., $K=7$.

For our proposed WD-based transfer learning approach, we implement it using four different machine learning models. Three of these—feedforward ANN, XGBoost, and thinned twinGP—were introduced earlier. In this section, we add a fourth model, the thinned scaled Vecchia (SV) method~\citep{chokhachian2025}, another competitive Gaussian process method designed for handling autocorrelated data. As shown in \citet{chokhachian2025}, thinnedSV runs more slowly than thinned twinGP but often delivers better performance. To disentangle the effects of model choice from the WD metric itself, we also develop geographic and pooled variants of ANN and thinned twinGP, as well as a WD-based variant of XGBoost.

\begin{table}[htbp]
\centering \footnotesize
\caption{Leave one turbine out experiment. The ``2017 RMSE'' is when models were trained on 2017 and tested on 2017 and the ``2018 RMSE'' is when models were trained on 2017 but tested on 2018. Boldface indicates the smallest RMSE in the respective testing year. Gaps are computed relative to the best method in the proposed WD Transfer section.}
\label{tab:rmse_loto}
\begin{tabular}{lcccc}
\toprule
\textbf{Method} & \multicolumn{2}{c}{\textbf{Average 2017 RMSE}} & \multicolumn{2}{c}{\textbf{Average 2018 RMSE}} \\
\cmidrule(lr){2-3} \cmidrule(lr){4-5}
 & \textbf{RMSE} & \textbf{Gap (\%)} & \textbf{RMSE} & \textbf{Gap (\%)} \\
\midrule
\multicolumn{5}{l}{\textit{The proposed WD Transfer ($K=7$)}} \\
\midrule
thinnedSV      & \textbf{3.24} & 0.00  & \textbf{3.43} & 0.00  \\
ANN            & 3.30          & 1.85  & 3.50          & 2.04  \\
XGBoost        & 3.32          & 2.47  & 3.56          & 3.79  \\
thinned twinGP & 3.34          & 3.09  & 3.58          & 4.37  \\
\midrule
\multicolumn{5}{l}{\textit{Pooled training (all turbines)}} \\
\midrule
STGP           & 3.73          & 15.12 & 3.97          & 15.74 \\
ANN            & 3.86          & 19.14 & 4.03          & 17.49 \\
thinned twinGP & 3.86          & 19.14 & 4.04          & 17.78 \\
XGBoost        & 3.89          & 20.06 & 4.09          & 19.24 \\
GNN            & 4.07          & 25.62 & 4.07          & 18.66 \\
BHM            & 4.03          & 24.38 & 4.15          & 20.99 \\
Binning        & 4.57          & 41.05 & 4.62          & 34.69 \\
\midrule
\multicolumn{5}{l}{\textit{Geographic-neighbor transfer ($K=7$)}} \\
\midrule
ANN            & 3.36          & 3.70  & 3.57          & 4.08  \\
thinned twinGP & 3.43          & 5.86  & 3.66          & 6.71  \\
Random Forest  & 3.60          & 11.11 & 3.84          & 11.95 \\
XGBoost        & 3.64          & 12.35 & 3.87          & 12.83 \\
SVR            & 3.60          & 11.11 & 3.93          & 14.58 \\
\bottomrule
\end{tabular}
\end{table}

Table~\ref{tab:rmse_loto} presents the results of the leave-one-turbine-out (LOTO) experiment. Within the proposed WD-based transfer group, thinnedSV emerges as the leading approach, holding a modest edge over the other three choices, with ANN trailing by 1--2\%, XGBoost trailing by 2--3\%, and thinned twinGP trailing by 3--4\%.

When all turbines are pooled into the training set, predictive RMSE rises considerably. The pooled models fall 15--41\% behind the leading WD-based method. Among these, STGP, which relies on terrain-based transfer, outperforms the rest by more than 3 percentage points. BHM, despite also being terrain-aware, fails to deliver competitive results. For ANN, switching from pooling to WD transfer yields a 15--17\% reduction in error; the corresponding margin is 13--16\% for thinned twinGP and 15--18\% for XGBoost.

In the geographic-neighbor transfer group, ANN and thinned twinGP perform better than others. They trail the leading WD-transfer thinnedSV method by 4--7\%, but when compared against their own counterparts in the WD group, the gap narrows to about 2\% for ANN and 2--3\% for thinned twinGP. Geographically transferred XGBoost falls 9--10\% behind its counterpart in WD. The remaining methods in this group fall further behind, by 11--15\%.

\subsection{Method Comparison on DFP Split}

Studying cross-farm power curve transfer, in a way that closely mimics the farm-planning scenario illustrated earlier in Figure~\ref{fig:fig1}, ideally requires data with two properties. First, all turbines should share the same make, model, and service duration, so that comparisons are not confounded by manufacturing or operational differences. Second, the data should span more than one farm, allowing a model to be trained on one farm and tested on another. Such homogeneous multi-farm datasets are rarely available publicly, posing a fundamental benchmarking challenge.

To work around this limitation, we exploit a unique feature of our single wind farm. As shown in Figure~\ref{fig:fig5}, a cluster of turbines (\#38–\#44) sits in the lower-left corner of the farm, geographically isolated from the rest by more than 2.2~km. We treat this cluster as the planning (testing) turbines and the remaining turbines (\#1–\#37 and \#45–\#66) as the existing (training) farm. This spatial separation lets us treat the two groups as if they belonged to two distinct farms, an arrangement we refer to as the distant farm planning (DFP) experiment. We train models on 2017 data from the training turbines (\#1–\#37 and \#45–\#66) and test them on 2018 data from the isolated cluster of turbines (\#38–\#44).  We acknowledge that the spatial separation is not great enough to be like two real wind farms, but hope that the analysis sheds interesting insights on the transfer learning problem at hand.

Since both WD transfer and geographic-neighbor transfer substantially outperformed pooled methods in the LOTO experiment, our primary goal in designing the DFP test is to examine whether these LOTO findings hold in a realistic planning setting. To this end, we compare geographic-neighbor transfer methods against WD transfer using DFP design.

\begin{table}[htbp]
\centering \footnotesize
\caption{Distant farm planning experiment. The ``2018 RMSE'' is when models 
were trained on 2017 but tested on 2018. This 2018 RMSE is averaged on 7 
testing turbines. Boldface indicates the smallest RMSE in the respective 
testing year. Gaps are computed relative to the best method in the proposed 
WD transfer section.}
\label{tab:rmse_dfp}
\begin{tabular}{lcc}
\toprule
\textbf{Method} & \multicolumn{2}{c}{\textbf{2018 RMSE}} \\
\cmidrule(lr){2-3}
 & \textbf{RMSE} & \textbf{Gap (\%)} \\
\midrule
\multicolumn{3}{l}{\textit{The proposed WD transfer ($K=7$)}} \\
\midrule
thinnedSV      & \textbf{4.49} & 0.00  \\
thinned twinGP & 4.54          & 1.11  \\
XGBoost        & 4.54          & 1.11  \\
ANN            & 4.55          & 1.34  \\
\midrule
\multicolumn{3}{l}{\textit{Geographic-neighbor transfer ($K=7$)}} \\
\midrule
thinned twinGP & 4.94          & 10.02 \\
Random Forest  & 4.96          & 10.47 \\
ANN            & 4.99          & 11.14 \\
XGBoost        & 5.00          & 11.36 \\
SVR            & 5.05          & 12.47 \\
\bottomrule
\end{tabular}
\end{table}

Table~\ref{tab:rmse_dfp} presents the predictive error of the different methods under the DFP experiment. The four methods in the proposed WD transfer group achieve the best performance, with only about a 1\% difference among them, compared with a larger gap of 2--4\% observed under the LOTO experiment. ThinnedSV achieves the best performance under the DFP setting, while the remaining three methods all fall short by roughly 1\%.

The geographic distance-based transfer methods perform noticeably worse, with gaps of 10--13\% relative to thinnedSV, highlighting that spatial proximity alone is an unreliable proxy for transferring power curve models from one farm to another farm. Note that ANN, thinned twinGP, and XGBoost are in both the proposed WD group and the geographic-neighbor group, and the WD-based versions of these methods consistently outperform their geographic-transfer counterparts by 9–-10\%. 

Compared with the LOTO experiments, the gap between the WD version and geographic-neighbor version remains roughly the same for XGBoost but the gap grows substantially for thinned twinGP and ANN. We believe that the change in performance is a result of using a separated test cluster. In LOTO, the left-out test turbine is still rather close to some of the training turbines.  In the entire farm, there is not a single turbine is well isolated from the rest.  In DFP, the test turbine has a minimal distance away from the training turbine (about 2 km).  Even with such a small value in minimal separation, the geographic-neighbor version of these methods has already shown a considerable performance deterioration. One can only imagine how much worse the geographic-neighbor version of these methods may become when the separation distance increases to tens or hundreds of miles in between.

 \subsection{Geographic vs. Distributional Source Selection}
\begin{figure}[htbp]
\centering
\includegraphics[width=0.49\linewidth]{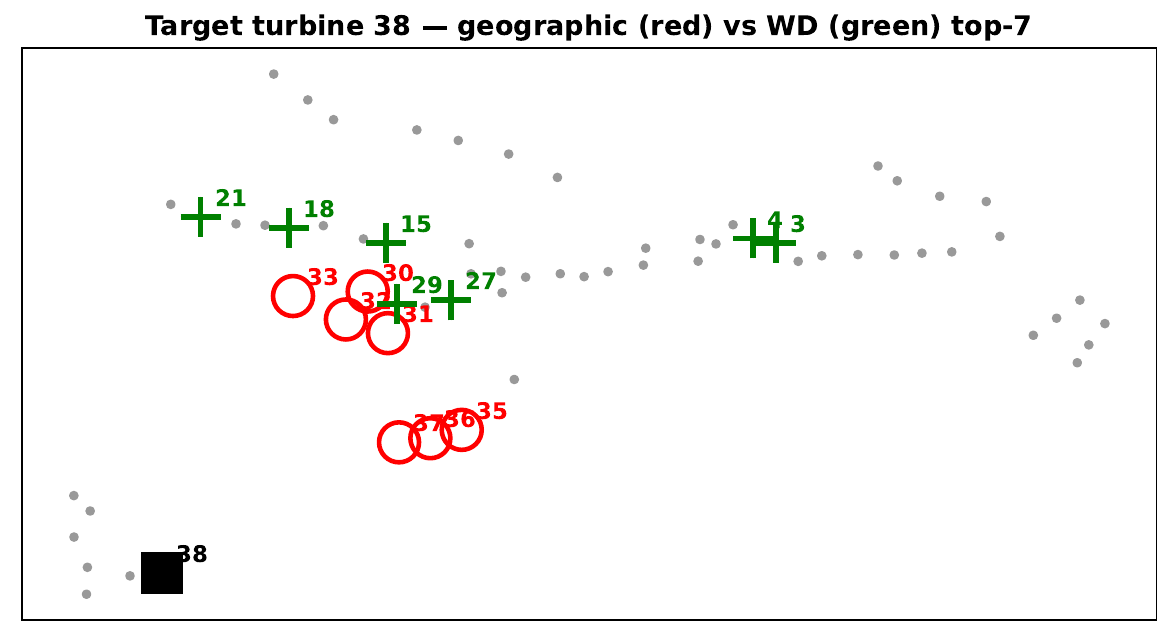}
\hfill
\includegraphics[width=0.49\linewidth]{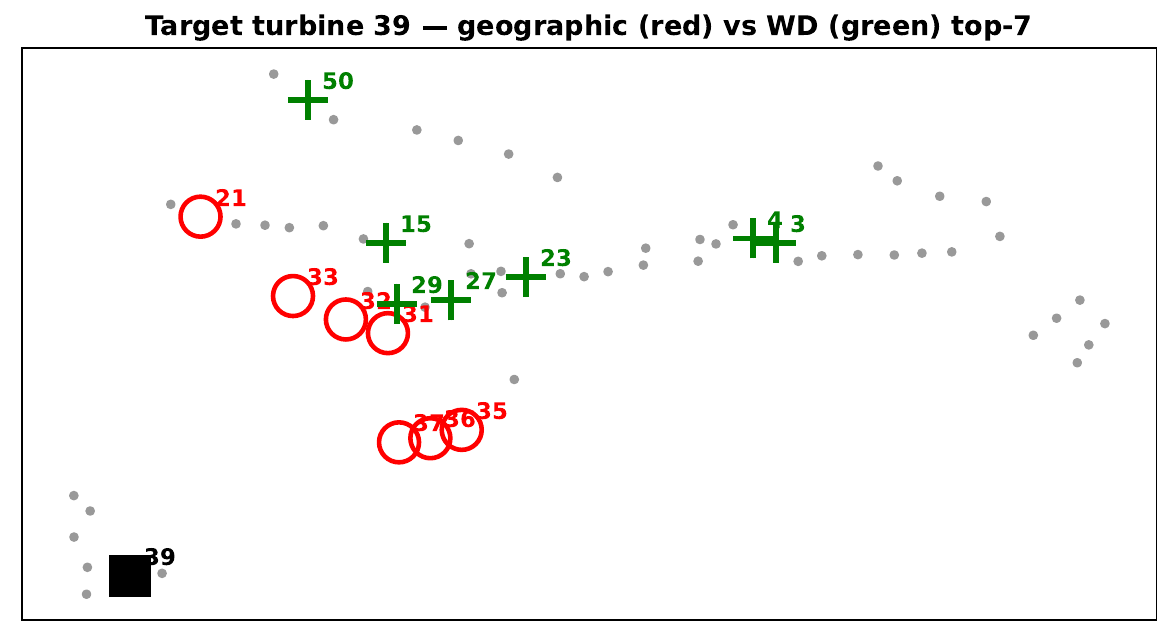}
\caption{For target turbines \# 38 (in the left plot) and \# 39 (in the right plot), the seven matched source turbines selected by the WD criterion are marked with a plus sign, whereas the seven geographically nearest neighbours are indicated by empty circles.}
\label{fig:WKSvsKNN}
\end{figure}

Geographic proximity, while intuitive as a source selection criterion, does not reliably translate into environmental or terrain similarity. Models trained on geographically proximal source turbines may fail to capture the inflow conditions most relevant to the target turbine. The comparisons in the two preceding sections demonstrate numerically the advantage of using the proposed weighted dissimilarity metrics for source selection.  Here we present an illustration, presented in Figure~\ref{fig:WKSvsKNN}, for making this point more intuitive.

Figure~\ref{fig:WKSvsKNN} illustrates this discrepancy between the two selection philosophies in the DFP setting.  It selects two turbines, \#38 and \#39, in that outlying smaller cluster, and use each as the target turbine.  Then select the source turbines to match the respective target turbine by using the two selection criteria: $K=7$ turbines using the weighted dissimilarity and $K=7$ turbines nearest to the target turbine. In each plot, the nearest neighbor selections are marked by empty circles, whereas those selected by WD are marked by a plus sign.  In these two cases, there is no overlap between the 14 turbines selected by the two criteria.  Several WD-selected turbines are located at a considerable distance from the target. 

This divergence between the two selection outcomes reflects the complex aerodynamics and terrain-driven processes that govern turbine inflow. Upstream wakes, terrain-induced flow acceleration and deceleration, and local features such as ridges and slopes can cause physically adjacent turbines to experience fundamentally different wind regimes and inflow conditions, while turbines that are farther apart but may exhibit more similar and comparable inflow distributions. The WD criterion captures these distributional similarities directly from the observed environmental covariates like wind speed, temperature, and terrain, whereas physical distance alone does not guarantee to deliver the same reliability in source matching and selection. As a result, WD-based source matching supply more informative training data for building a target turbine's power curve, leading to the consistent performance advantage of the proposed transfer learning framework over other alternative methods across both the LOTO and DFP experimental settings.

While Figure~\ref{fig:WKSvsKNN} presents only two target turbines, the observed pattern is not isolated. For the seven DFP test turbines, each requires $K = 7$ source turbines, so that there are $7 \times 7 = 49$ source turbines in total to be selected. Of these 49 turbines, only 5 of them are a common choice by both criteria, yielding an overlap rate of $5/49 \approx 10\%$. The remaining 44 selections are distinct between the choices made by the two criteria. 

In the LOTO experiment setting, the candidate pool is bigger with 462 source turbines to be selected.  The overlap among the selections by the two criteria rises to $29\%$, which still indicates substantial divergence. These numbers underscore that geographic proximity and environmental similarity identify fundamentally different source sets, and it matters a lot to use the correct and effective source selection criterion when it comes to building a cross-farm power curve.

\section{Conclusion}
\label{sec:conclusion}

This paper addresses cross-farm power curve modeling through a domain adaptation lens built on principled source selection. Source turbines are selected using a weighted dissimilarity metric that combines environmental and terrain features, with feature importance weights derived from accumulated local effects. The weights are representative of the importance of each feature in explaining the variance of power output. When used with a predictive ensemble strategy, a fast Gaussian process model delivers a cross-farm predictive performance better than all other alternatives by a comfortable margin. 

We would like to reiterate two important messages from this study. The first is the importance of source selection, which is a central theme in transfer learning and domain adaptation. Even for turbines of the same model and make, using all turbines indiscriminately for training a cross-farm power curve is not an effective strategy. Not all the turbines are subject to the same environmental conditions as the target turbine would experience.  When using all turbines, one is effectively using the \emph{average turbine} as the representation. Naturally, a better strategy is to find the subset of turbines that resembles the target turbine most closely.  

The next message is how to define the resemblance (i.e., similarity) and then use that for selecting the subset of source turbines that matters a lot.  The traditional wisdom in the wind energy literature relies on geographic distances, layout, or recently terrain for selection; those do not produce effective source selection. The proposed weighted dissimilarity metric, which combines both algebraic absolute difference and probabilistic distributional difference, wins the battle in the current study.  

Whether the proposed WD-based selection criterion should always be used for future cross-farm power curve developments remains to be seen. We believe that we have presented a strong case, but we admit that the proposed approach has only been tested on a single wind farm. It would have made a stronger case if a test could be conducted on two physically distinct wind farms with a realistic distance in between. It will also lend additional credibility to the proposed approach when more testing cases are conducted if they affirm the main message in this study. It is very well possible that certain shortcomings of the proposed approach are discovered in future testing, but that would give rise to opportunities for improving the method.


\section*{Supplementary Material}
\subsection*{Dataset and Computer Code}
\label{Datasets}
The dataset and computer code to reproduce all the results are publicly available at the \href{https://github.com/anonymous-stgp/Domain-Adapted-PC.git}{Domain-Adapted-PC} GitHub page. All experiments are implemented and executed on the Georgia Tech Partnership for an Advanced Computing Environment (PACE) high-performance computing cluster. Jobs are submitted through the Inferno service using CPU-only Intel processor nodes, with each job allocated one node and eight cores. 
    
\section*{Declaration of competing interest}
The authors declare that they have no known competing financial interests or personal 
relationships that could have appeared to influence the work reported in this paper.




\section*{Funding}

YD’s research was partially supported by National Science Foundation (NSF) Grant CNS--2328395.

\vspace{6 pt}

\spacingset{1}
\bibliographystyle{apalike}

\bibliography{bibio}

@article{chokhachian2026,
author = {Ahmadreza Chokhachian and V. Roshan Joseph and Yu Ding},
title = {Spatio-Temporal {G}aussian Process for Building Terrain-Incorporating Wind Power Curves},
journal = {Technometrics},
volume = {0},
pages = {1--26},
year = {2026}
}

@article{Wang2023,
  title={A novel data-driven deep learning approach for wind turbine power curve modeling},
  author={Y. Wang AND X. Duan AND R. Zou AND F. Zhang AND Y. Li AND Q. Hu},
  journal={Energy},
  volume={270},
  number={1},
  pages={126908},
  year={2023},
}

@article{rizzo2016,
  title={Energy distance},
  author={Rizzo, Maria L and Sz{\'e}kely, G{\'a}bor J},
  journal={Wiley Interdisciplinary Reviews: Computational statistics},
  volume={8},
  number={1},
  pages={27--38},
  year={2016},
}

@article{chizat2020,
  title={Faster Wasserstein distance estimation with the Sinkhorn divergence},
  author={Chizat, Lenaic and Roussillon, Pierre and L{\'e}ger, Flavien and Vialard, Fran{\c{c}}ois-Xavier and Peyr{\'e}, Gabriel},
  journal={Advances in neural information processing systems},
  volume={33},
  pages={2257--2269},
  year={2020}
}

@ARTICLE{Kusiak2009,
  AUTHOR = "A. Kusiak and H. Zheng and Z. Song",
	YEAR = "2009",
  TITLE = "On-line monitoring of power curves",
  JOURNAL = "Renewable Energy",
	VOLUME = "34",
  PAGES = "1487--1493"}

@article{Pandit2020,
  title={{SCADA} Data-Based Support Vector Machine Wind Turbine Power Curve Uncertainty Estimation and Its Comparative Studies},
  author={R. K. Pandit AND A. J. Kolios},
  journal={Applied Sciences},
  volume={10},
  number={23},
  pages={8685},
  year={2020},
}

@article{Yesilbudak2013,
  title={A new approach to very short term wind speed prediction using k-nearest neighbor classification},
  author={M. Yesilbudak AND S. Sagiroglu AND I. Colak},
  journal={Energy Conversion and Management},
  volume={69},
  pages={77--86},
  year={2013},
}

@article{Ding2022,
  author  = {Ding, Yu and Barber, Sarah and Hammer, Florian},
  title   = {Data-driven wind turbine performance assessment and quantification using field measurements},
  journal = {Frontiers in Energy Research, section Wind Energy},
  volume  = {10},
  pages   = {1050342},
  year    = {2022}
}

@book{Gu2013,
  author    = {Gu, C.},
  title     = {Smoothing Spline ANOVA Models},
  publisher = {Springer},
  address   = {New York, NY, USA},
  edition   = {2},
  year      = {2013}
}

@article{Lee2015b,
  author  = {Lee, G. and Ding, Y. and Xie, L. and Genton, Marc G.},
  title   = {Kernel Plus Method for Quantifying Wind Turbine Upgrades},
  journal = {Wind Energy},
  volume  = {18},
  pages   = {1207--1219},
  year    = {2015}
}

@article{Lee2015a,
  author  = {Lee, G. and Ding, Y. and Genton, Marc G. and Xie, L.},
  title   = {Power Curve Estimation with Multivariate Environmental Factors for Inland and Offshore Wind Farms},
  journal = {Journal of the American Statistical Association},
  volume  = {110},
  pages   = {56--67},
  year    = {2015},
  doi     = {10.1080/01621459.2014.977385}
}

@article{chokhachian2025,
  title        = {Fast {G}aussian Process Approximations for Autocorrelated Data},
  author       = {Ahmadreza Chokhachian and Matthias Katzfuss and Yu Ding},
  journal      = {INFORMS Journal on Data Science},
  year         = {2026},
  volume       = {online published},
  pages       = {1--17}
}

@article{Hammer2022,
year = {2022},
volume = {2151},
number = {1},
pages = {012006},
author = {Hammer, Florian and Barber, Sarah},
title = {Transferability of site-dependent wind turbine performance predictions using machine learning},
journal = {Journal of Physics: Conference Series}
}

@article{li2022b,
  title={Transfer strategy for power output estimation of wind farm at planning stage based on a {SVR} Model},
  author={Li, Zihao and Sun, Wei and Xiang, Yue and Harrison, Gareth P},
  journal={CSEE Journal of Power and Energy Systems},
  volume={9},
  number={4},
  pages={1460--1471},
  year={2022},
  publisher={CSEE}
}

@article{Daenens2025,
year = {2025},
volume = {3016},
pages = {012021},
author = {Daenens, Simon and Verstraeten, Timothy and Daems, Pieter-Jan and Nowé, Ann and Helsen, Jan},
title = {Power Prediction in Offshore Wind Farms using Transferable Multi-Task Graph Neural Networks},
journal = {Journal of Physics: Conference Series}
}

@article{Arrieta2024,
title = {Spatially transferable machine learning wind power prediction models: v{-}logit random forests},
journal = {Renewable Energy},
volume = {223},
pages = {120066},
year = {2024},
author = {Mario Arrieta-Prieto and Kristen R. Schell}
}

@inproceedings{Barber2024,
  author    = {S. Barber and Y. Ding},
  title     = {Improving data sharing in practice – power curve benchmarking case study},
  booktitle = {Proceedings of the 2024 WindEurope Annual Event, Bilbao, Spain, March 20-22},
  year      = {2024}
}

@article{PanYang2010,
  author  = {Pan, Sinno Jialin and Yang, Qiang},
  title   = {A survey on transfer learning},
  journal = {IEEE Transactions on Knowledge and Data Engineering},
  year    = {2010},
  volume  = {22},
  pages   = {1345--1359}
}

@article{Prakash2023,
  author  = {Prakash, Abhinav and  Tuo, Rui and Ding, Yu},
  title   = {The temporal overfitting problem with applications in wind power curve modeling},
  journal = {Technometrics},
  volume  = {65},
  pages   = {70--82},
  year    = {2023}
}

@article{Prakash2024,
  author  = { Prakash, Abhinav and  Lee, Se Yoon and Liu, Xin and Liu, Lei and  Mallick, Bani K and Ding, Yu},
  title   = {A {B}ayesian hierarchical model to understand the effect of terrain on wind turbine power curves},
  journal = {IEEE Transactions on Sustainable Energy},
  volume  = {15},
  pages   = {1127--1137},
  year    = {2024}
}

@article{Apley2016,
  title={Visualizing the effects of predictor variables in black box supervised learning models},
  author={Apley, Daniel W. and Zhu, Jingyu},
  journal={Journal of the Royal Statistical Society: Series B},
  volume={82},
  pages={1059--1086},
  year={2020}
}

@book{Ding2019,
  author    = {Yu Ding},
  title     = {Data Science for Wind Energy},
  publisher = {Chapman \& Hall},
  address   = {Boca Raton, FL, USA},
  year      = {2019}
}

@Manual{XGB,
    title = {Xgboost: Extreme Gradient Boosting},
    author = {Tianqi Chen and Tong He and Michael Benesty and Vadim
      Khotilovich and Yuan Tang and Hyunsu Cho and Kailong Chen and
      Rory Mitchell and Ignacio Cano and Tianyi Zhou and Mu Li and
      Junyuan Xie and Min Lin and Yifeng Geng and Yutian Li and Jiaming
      Yuan and David Cortes},
    year = {2025},
    note = {R package version 3.0.2.1 available at \url{https://github.com/dmlc/xgboost}},

}

@book{Kumar2020,
title = "{DSWE: Data Science for Wind Energy, \texttt{R} Package}",
author = {Kumar, N. and Prakash, A. and Ding, Y},
year = {2020},
publisher= {The Comprehensive \texttt{R} Archive Network (\texttt{CRAN}) https://cran.r-project.org/web/packages/DSWE/index.html}
}

@book{Kumar2022,
title = "{DSWE: Data Science for Wind Energy, \texttt{Python} Package}",
author = {Kumar, P. and Prakash, A and Ding, Y},
year = {2022},
publisher = {The \texttt{Python} Package Index (\texttt{PyPI}) https://pypi.org/project/dswe/}
}

@book{IEC2005,
  author       = {IEC},
  title        = {{IEC TS 61400-12-1 Ed. 1: Power Performance Measurements of Electricity-Producing Wind Turbines}},
  publisher    = {International Electrotechnical Commission},
  address      = {Geneva, Switzerland},
  year         = {2005},
  note         = {Technical Specification}
}

@book{IEC2013,
  author       = {IEC},
  title        = {{IEC 61400-12-2 Ed. 1.0: Wind Turbines—Part 12-2: Power Performance of Electricity-Producing Wind Turbines Based on Nacelle Anemometry}},
  publisher    = {International Electrotechnical Commission},
  address      = {Geneva, Switzerland},
  year         = {2013},
  note         = {International Standard}
}

\end{document}